\documentclass[letterpaper, 10 pt, conference]{ieeeconf}  

\IEEEoverridecommandlockouts                              

\usepackage{color}

\usepackage{times}
\usepackage{epsfig}
\usepackage{graphicx}
\usepackage{tabularx}
\usepackage{amsmath}
\usepackage{amssymb}
\usepackage{bbm}
\usepackage{svg}
\usepackage[misc]{ifsym}
\usepackage[export]{adjustbox}

\usepackage[ruled,vlined,linesnumbered]{algorithm2e}
\usepackage{graphicx} 

\usepackage{tikz}
\usepackage{verbatim}
\usepackage{booktabs}
\usepackage{multirow}
\usepackage{makecell}
\usepackage{authblk}
\usepackage{hyphenat} 

\usepackage{pgf}
\usepackage{algpseudocode}
\usetikzlibrary{arrows,automata}
\makeatletter
\let\NAT@parse\undefined
\makeatother
\usepackage[colorlinks]{hyperref}  

\title{
Needs-driven Heterogeneous Multi-Robot Cooperation in Rescue Missions}

\author{Qin Yang}
\author{Ramviyas Parasuraman$^*$\thanks{* Corresponding author email: ramviyas@uga.edu}}
\affil{HeRo Lab, Department of Computer Science, University of Georgia, Athens, GA 30605, USA}

\DeclareMathOperator*{\argmax}{arg\,max}

\begin{document}
\bstctlcite{IEEEexample:BSTcontrol}

\newtheorem{definition}{Definition}
\newtheorem{theorem}{Theorem}
\newtheorem{lemma}{Lemma}
\newtheorem{proposition}{Proposition}
\newtheorem{property}{Property}
\newtheorem{observation}{Observation}
\newtheorem{corollary}{Corollary}

\maketitle
\thispagestyle{empty}
\pagestyle{empty}

\begin{abstract}

This paper focuses on the teaming aspects and the role of heterogeneity in a multi-robot system applied to robot-aided urban search and rescue (USAR) missions. We propose a needs-driven multi-robot cooperation mechanism represented through a Behavior Tree structure and evaluate the system's performance in terms of the group utility and energy cost to achieve the rescue mission in a limited time. From the theoretical analysis, we prove that the needs-driven cooperation in a heterogeneous robot system enables higher group utility than a homogeneous robot system. We also perform simulation experiments to verify the proposed needs-driven collaboration and show that the heterogeneous multi-robot cooperation can achieve better performance and increase system robustness by reducing uncertainty in task execution. Finally, we discuss the application to human-robot teaming.

\end{abstract}

\section{Introduction}
\begin{figure}[tbp]
\centering
\includegraphics[width=0.48\textwidth]{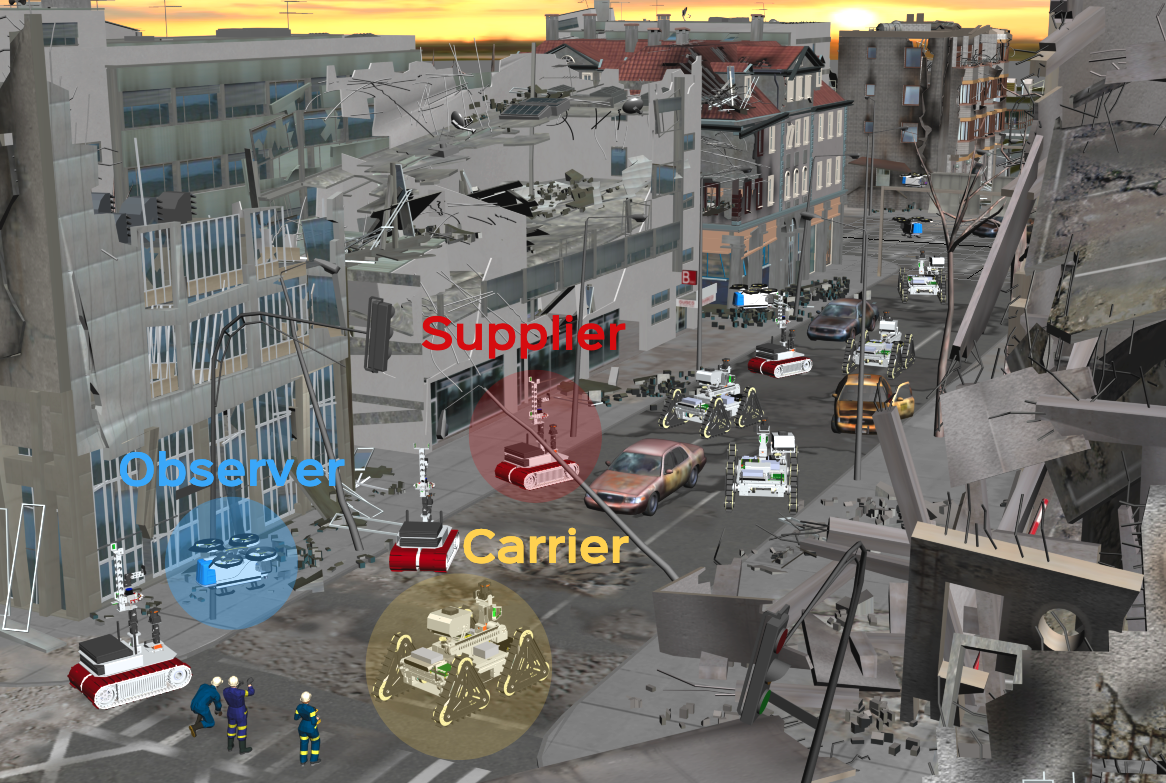}
\caption{Illustration of an integrated team of robots (UGVs + UAVs) and human cooperatively working together in a post-earthquake rescue mission.}
\label{fig: overview}
\end{figure}

Rescue missions can be regarded as life-saving, delivering valuable properties, and tackling necessary facilities in disaster or emergency scenarios, including complex, hazardous, uncertain, unstructured, dynamical changing, and adversarial environments. Multi-Robot System (MRS) working in such situations requires rapid response, high adaptation, and strong robustness, reducing the losses in the post-disaster scenarios. Research in robot-aided USAR aims to increase the mission success rate, improve execution efficiency, and minimize system cost during the rescue missions. 
Fig. \ref{fig: overview} illustrates an example real-world use-case of MRS in a post-earthquake scenario, where we represent teams of three different robot types - \textit{Carrier}, \textit{Supplier}, and \textit{Observer} - aiding the first responders in close collaboration.

Disasters are defined as discrete meteorological, geological, or man-made events that exceed local resources to respond and contain \cite{Murphy2016}.
From the robot's needs \cite{yang2020hierarchical} and motivations perspective, we can classify Adversaries into two general categories in disaster or adversarial environments. One is Intentional (such as enemy or intelligent opponent agent, which consciously and actively impairs the MAS needs and capabilities), and the other is Unintentional (like obstacles and weather, which unaware and passively threaten MAS abilities) adversary.\cite{yang2020gut}. 
We are specifically interested in the MRS collective tackling the \textit{Unintentional Adversary} in hazardous and disaster scenarios. So the environment models for rescue missions are grounded in two different aspects: individual perception and data sharing across the robots.

Considering individual perception, we emphasize cooperation among heterogeneous groups of robots \cite{stone2000multiagent}. Each robot class might have different sensors and capabilities to perceive and interact with the environment and corresponding actuators to execute their action. Individual robots present their observations from different angles describing the partial part in the global map.
Regarding system data sharing, each robot in the current group needs to update its situation awareness from other group members' information. It can not only help in collectively building a global map \cite{rizk2019cooperative} but also be a foundation for communication between the agents to achieve consensus \cite{parasuraman2018multipoint} or \textit{Negotiation} \cite{yang2020hierarchical}. 

It is essential to understand how to combine a team of mobile robots to achieve a successful search and rescue mission, especially from a heterogeneity point of view and through needs-driven cooperation among robots. Therefore, in this paper, we analyze the association between heterogeneous robots with different capabilities and needs for MRS collaboration and teamwork. Specifically, we make the following contributions in this paper:
\begin{itemize}
    \item We generalize the rescue mission's problem using different groups of robots, such as Carrier, Supplier, and Observer. We formalize the multi-robot cooperation through robot needs hierarchy encoded in a Behavior Tree \cite{colledanchise2018behavior} structure;
    \item We theoretically analyze the rescue robot teaming from two perspectives: \textit{Utility} achieved by the robot group and \textit{Energy} consumed by the group.  
    \item We verify the theoretical results through simulations with different teams of homogeneous and heterogeneous robots deployed to a rescue mission.
\end{itemize}

\begin{figure}[tbp]
\centering
\includegraphics[width=0.5\textwidth]{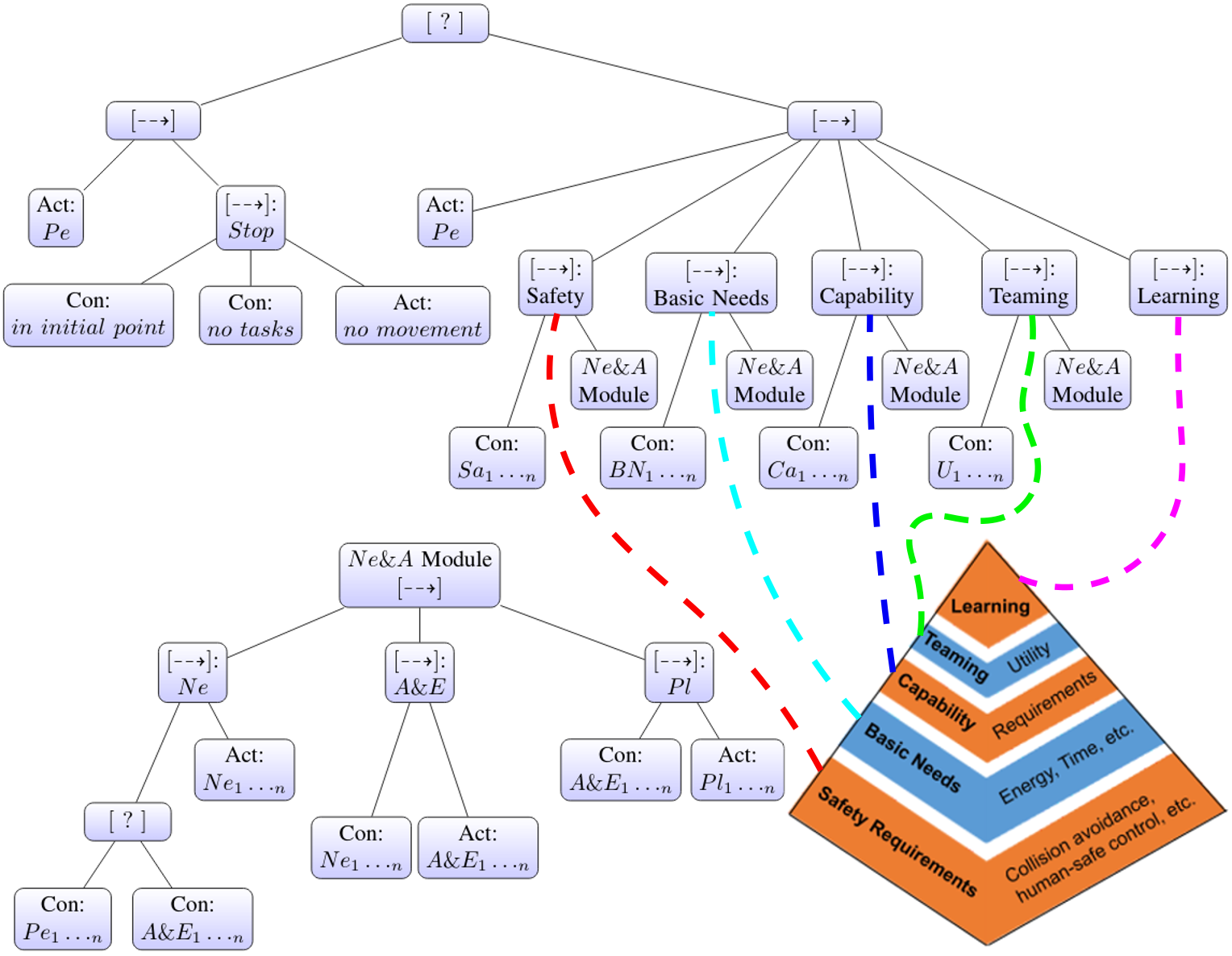}
\caption{Behavior Tree representing hierarchy of robot needs at every robot. $[?]$ - Selector Node, $[\dashrightarrow]$ - Sequence Node, $Con$ - Conditions, $Act$ - Actions, $Pe$ - Perception, $Sa$ - Safety, $BN$ - Basic Needs, $Ca$ - Capability, $U$ - Utility, $Pl$ - Plan, $Ne$ - Negotiation, $A\&E$ - Agreement and Execution.}
\label{fig: bt}
\end{figure}

\section{Related Works and Projects}
\label{sec:relatedwork}

An intelligent agent is a physical (robot) or virtual (software program) entity that can autonomously perform actions on an environment while perceiving this environment to accomplish a goal \cite{russell2002artificial}. 
Cooperation in multiple intelligent agents (robots) working in a disaster environment is a interesting and challenging problem \cite{Murphy2016, jorge2019survey}. Most research focus on the problems of environmental monitoring \cite{byrne2012study, bayat2017environmental, marques2015critical}, structure inspection \cite{moud2018current, lattanzi2017review}, navigation and control \cite{fossen2000survey, ashrafiuon2010review, campbell2012review,luo2019multi} and higher-level autonomy \cite{schiaretti2017survey, thompson2019review}. 
Also, there are various advancements in rescue robotics through the development of heterogeneous robot teaming methods in disaster response scenarios \cite{kruijff2015tradr, 6719323, marconi2013ground,nourbakhsh2005human}, disaster detection \cite{liu2016usv, fornai2016autonomous}, disaster monitoring \cite{vasilijevic2017coordinated, guerrero2016multirobot}, target tracking \cite{rathour2015sea, fahad2017evaluation}, victims detection \cite{cardona2019robot}, and reinforcement learning based semi-autonomous controller for urban search and rescue missions \cite{magid2020artificial}. 

Considering grouping robots with various capabilities cooperating to pursue specific goals (rescue missions), less literature study the integration of organizing agents' behaviors, solving the conflicts, optimizing system utility, and boosting system adaptability and robustness for the entire group \cite{yang2018grand, rizk2019cooperative}. On the other hand, there is little research done from the agent's needs perspective studying individual interaction and behaviors for system performance (group utility) and global actions in MRS, especially in disaster robotics \cite{mrs2019,yang2020hierarchical,geihs2020engineering}.

To address those gaps, we build upon our work in \cite{yang2020hierarchical, mrs2019}, where we represent complex relationships between different types of robots through their immediate needs and motivations. It helps the system to balance and optimize the utilities between the individual and the whole group. 
We encode the individual robot needs hierarchy in the robot automated planner represented through a Behavior Tree structure \cite{colledanchise2018behavior, colledanchise2018learning}. Then we analyze the MRS group performance by theoretically deriving and comparing the group utility and their energy cost applied to a USAR mission. Also, the \textit{Human-Robot} mixed teaming by combining human and robot needs will benefit from the study. More importantly, it can improve system adaptability and solve more complex tasks. On the other hand, it helps individual self-upgrade and self-evolution of the whole system through \textit{Adaptation Learning} from interaction and experience between robots and humans.

\section{Needs-driven Model for Robot Cooperation}
\label{sec:needsmodel}

In nature, from cell to human, all intelligent agents represent different kinds of hierarchical needs such as the low-level physiological needs (food and water) in microbe and animal; the high-level needs self-actualization (creative activities) in human being \cite{maslow1943theory}. Simultaneously, intelligent agents can cooperate or against each other based on their specific needs. As an artificial intelligence agent -- robot, to organize its behaviors and actions, we introduced the needs hierarchy of robots in \cite{yang2020hierarchical} to help MRS build cooperative strategies considering their individual and common needs. Specifically, the robots possess the following order of needs hierarchy: Safety needs (avoid collisions, safe environment, etc.); Basic needs (Energy, time, mobility, etc.); Capability needs (task-specific such as carry or supply resources); Teaming needs (enhancing group utility and group survival); and Learning needs (self-upgrade and evolution).

Since the robot needs to rescue the victims from the disaster or cooperate with people to fulfill rescue missions together, the robot's lowest level needs should guarantee human safety and security. This kind of condition reflex or self-reactive behavior in robots can be represented as fundamental control issues like collision avoidance. After satisfying the safety needs, the robot requires enough basic needs, like battery, oil, to support executing relative operations. Then, by comparing their capabilities and the task requirements, they will select how to cooperate maximizing the success rate in rescue missions, and optimize or sub-optimize individual and system \textit{utility}.

To fulfill a high level needs satisfying individual or group's \textit{expectation utilities} \cite{yang2020gut}, different categories of robots consider working as one or several teams to maximize corresponding utilities or rewards efficiently. When assigned with new rescue tasks or encounter emergency events like some group members run out of battery, robots need to re-organize the group adapting to the current situation minimizing the cost and loss. Fig.~\ref{fig: bt} presents an individual robot hierarchy of needs encoded in the form of a state-of-the-art state-action planner called Behavior Trees \cite{colledanchise2018behavior}. 

In rescue missions, we consider the \textit{Group's Utility} as the number of lives (victims) or valuable properties saved and rescued as much as possible in a limited time. In the entire process, robots need to consider exploring the uncertain area, tackling the \textit{unintentional adversaries} like obstacles, wind, rain, and so on, repairing necessary facilities, treating injurers, carrying victims and properties to a safe place.

\section{Formalization and Evaluation}

This section first formalizes the rescue problem and uses mathematical approaches to prove our hypothesis that cooperation in heterogeneous robot system produce better performance in general than the cooperation limited to a homogeneous robot system, with rescue mission as an example application domain.

Consider the following example. Supposing a group of heterogeneous robots executes the search and rescue mission in a post-disaster scenario. The robot's categories can be generally classified as follows.

\begin{itemize}
    \item \textbf{Carrier:} Their main function is carrying injurers and valuable properties from hazardous areas to shelter.
    \item \textbf{Supplier:} Providing various resources for rescue missions such as medicine, food, repairing robots, rescue devices, communication support, and so forth.
    \item \textbf{Observer:} They are good at surveying and acquiring real-time and dynamical rescue information from the disaster environment. 
\end{itemize}

\subsection{Problem Statement}
\label{sec:problem}
As discussed in Sec. \ref{sec:needsmodel}, we assume that the number of \textit{Carrier}, \textit{Supplier} and \textit{Observer} are $x$, $y$ and $z$ ($x, y, z \in Z^+$), respectively. We define the individual capability space according to the robot needs model through the below equations.
\begin{eqnarray}
    & Carrier := C_C(v_c, com_c, sen_c, eng_c, res_c, cap_c); \label{carrier} \\
    & Supplier := C_S(v_s, com_s, sen_s, eng_s, res_s, cap_s); \label{supplier} \\
    & Observer := C_O(v_o, com_o, sen_o, eng_o, res_o, cap_o). \label{observer}
\end{eqnarray}
Here, 
\begin{itemize}
\item $v$ represents agent's max velocity;
\item $com$ and $sen$ represent the range of agent's communication and sensing separately;
\item $eng$ represents agent's energy level;
\item $res$ represents the amount of rescue resource which agent can provide;
\item $cap$ represents agent's the capacity level.
\end{itemize}

Since each type of robot specialize in different capability, we can assume Eqs. \eqref{com}, \eqref{sen}, \eqref{vel}, \eqref{eng}, \eqref{res}, \eqref{cap}  showing the dominance of each robot type (denoted with subscripts $c,s,o$ to represent carrier, supplier, and observer robots, respectively) in different capabilities in terms of sensing and communication ranges, energy level capacities, etc. 

\textit{Robot Safety Needs:}
\begin{eqnarray}
    & com_o \gg com_s \approx com_c; \label{com} \\
    & sen_o \gg sen_s \approx sen_c; \label{sen} \\
    & v_o > v_s \approx v_c; \label{vel}
\end{eqnarray}

\textit{Robot Basic Needs:}
\begin{eqnarray}
    eng_c > eng_s \gg eng_o \label{eng}
\end{eqnarray}

\textit{Robot Capabilities for Rescue Mission Requirement:}
\begin{eqnarray}
    & res_s \gg res_c > res_o; \label{res} \\
    & cap_c \gg cap_s > cap_o; \label{cap}
\end{eqnarray}

Supposing rescue mission $T$ has requirement space $TC = (C_1, C_2, ... , C_m),~m \in Z^+$, where $C_i$ represents different capabilities expected required to achieve a given global task and $m$ is the capacity of the  required to satisfy the tasks. We assume that the heterogeneous group capabilities for rescue mission requirements is $C_G = (..., C_{C_x}, ..., C_{S_y}, ..., C_{O_z})$ and group members' expected round-trips within $t\leq t_n$ is  $m(m_1, ..., m_k) ~k \in Z^+$, where $k = x+y+z$. $U$ and $t_n$ represent the rescue mission's \textit{Group Utility} and mission time restriction, respectively. Then, we can regard rescue problem as an optimization problem Eq. \eqref{rescue_mission_problem}, which means that in the limited time, fulfilling a rescue mission maximum its \textit{Expectation Utility} based on certain requirements.
\begin{equation}
\begin{split}
    & \underset{C_G}\argmax~~~ \mathop{\mathbb{E}(U(t_n, m \cdot C_G))} ; \\
    & subject~to~~~ \sum_{d=1}^{x} \sum_{e=1}^{y} \sum_{f=1}^{z} m \cdot C_G \geqslant TC,~d, e, f \in Z^+.
\label{rescue_mission_problem}
\end{split}
\end{equation}

To simplify our model, we just consider one specific rescue mission and $n$ identical obstacles distributed in an uncertain disaster environment randomly. The encountering obstacles times $X$ for each agent follow \textit{Poisson Distribution} Eq. \eqref{normal} and $\lambda$ represents as Eq. \eqref{lam} ($c$ and $sen$ are corresponding coefficient and area of sensing range).
\begin{eqnarray}
    && X \sim P(\lambda); \label{normal} \\
    && \lambda = \frac{cn}{sen} \label{lam}
\end{eqnarray}

And we assume that the average time and energy cost for individuals tackling each obstacle are $t_c$ and $e_c$, respectively. The distance between the central point of the initial group and rescue position is $l$. We also assume that agent energy costs mainly consist of traveling, tackling the obstacles, and fulfilling the rescue task \cite{parasuraman2012energy}. The traveling energy cost can be regarded as constant $e_t$, which is proportional to $l$.

Through Eqs. \eqref{normal} and \eqref{lam}, we can easily calculate the expectation of time encountering obstacles as Eq. \eqref{times}. Without considering obstacles, individuals coming to rescue position and returning to initial point energy cost are $\frac{2l}{v}$. Then, considering the obstacles, we estimate the expected time cost per round as Eq. \eqref{time}.
\begin{eqnarray}
    && \mathop{\mathbb{E}(X) = \sum_{i=0}^{+\infty}iP(X=i) = \lambda = \frac{cn}{sen}}; \label{times} \\
    && \mathop{\mathbb{E}(T) = \mathbb{E}(\frac{2l}{v} + 2t_cX)} = \frac{2l}{v} + \frac{2t_ccn}{sen} \label{time}
\end{eqnarray}

\subsection{Theoretical Evaluation}

In this section, we generally classify the rescue team as two different categories: \textit{Homogeneous} and \textit{Heterogeneous}.
A homogeneous robot system means the robots in that group are of the same type with same capabilities (Eq.~\eqref{cap}), whereas robots in a heterogeneous group will have different capabilities \cite{stone2000multiagent,twu2014measure}.
We assume that each agent's sensing range equal to its communication range for the sake of simplicity in analysis. Then, we use mathematical approaches to analyze and compare their performance as follows:

\paragraph{Homogeneous Cooperation}

In this scenario, we suppose that the number of \textit{Carrier}, \textit{Supplier} and \textit{Observer} are equal Eq. \eqref{num}. According to Eq. \eqref{time}, the time homogeneous group per round can be represented as Eq. \eqref{time1}.
\begin{eqnarray}
    && x = y = z = 3m,~~m \in Z^+; \label{num} \\
    && \mathop{\mathbb{E}(T_h) = \frac{2l}{v} + \frac{2t_c cn}{3m \times sen} = \lambda_h} \label{time1}
\end{eqnarray}

Considering rescue mission's tackling time equal to one unit time, the entire expectation rescue per round time is $\mathop{\mathbb{E}(T_h + 1)}$. Then, we can calculate $\mathop{\mathbb{E}(\frac{1}{T_h + 1})}$ as Eq. \eqref{time2}.
\begin{equation}
\begin{split}
    \mathop{\mathbb{E}(\frac{1}{T_h + 1})} & = \sum_{k=0}^{+\infty}\frac{1}{k+1}P(T_h = k) \\
    & = \sum_{k=0}^{+\infty}\frac{1}{k+1}\frac{\lambda_h^ke^{-\lambda_h}}{k!} \\
    & = \frac{1}{\lambda_h}\sum_{k=0}^{+\infty}\frac{\lambda_h^{k+1}e^{-\lambda_h}}{(k+1)!} \\
    & = \frac{1}{\lambda_h}\sum_{k=0}^{+\infty}P(T_h = k + 1) \\
    & = \frac{1}{\lambda_h}\sum_{k=1}^{+\infty}P(T_h = k) \\
    & = \frac{1 - P(T_h = 0)}{\lambda_h} = \frac{1 - e^{-\lambda_h}}{\lambda_h}
\label{time2}
\end{split}
\end{equation}

Finally, we estimate the expected number of rounds for this homogeneous group in the rescue mission with a limited time $t_n$ as Eq. \eqref{time3}.
\begin{eqnarray}
    \mathop{\mathbb{E}(\frac{t_n}{T_h+1}) = \frac{t_n(1 - e^{-\lambda_h})}{\lambda_h}} \label{time3}
\end{eqnarray}

We are supposing rescuing each agent cost one point supplement, energy, and space, respectively. 
We can estimate the sum of \textit{Expectation Utility} -- the amount of rescued agents $U_h$ in all rounds and the total energy cost $E_h$ in group of \textit{Carrier}, \textit{Supplier} and \textit{Observer} as Eq. \eqref{res_cso} and \eqref{eng_cso} respectively.

\textit{\small{a. Carrier/Supplier/Observer expectation amount of rescued agents}}
\begin{equation}
\begin{split}
    & \mathop{\mathbb{E}(U_{hc/s/o}) = \frac{t_n(1 - e^{-\lambda_{hc/s/o}})}{\lambda_{hc/s/o}}3m \times res_c/cap_s/cap_o}, \\
    & \lambda_{hc/s/o} = \frac{2l}{v_{c/s/o}} + \frac{2t_ccn}{3m \times sen_{c/s/o}}
\label{res_cso}
\end{split}
\end{equation}
\textit{\small{b. Carrier/Supplier/Observer expectation energy cost}}
\begin{equation}
\begin{split}
    \mathop{\mathbb{E}(E_{hc/s/o})} = 
    & e_t + \frac{2cn}{3m \times sen_{c/s/o}}e_c + \\
    & \frac{t_n(1 - e^{-\lambda_{hc/s/o}})}{\lambda_{hc/s/o}}3m \times res_c/cap_s/cap_o
\label{eng_cso}
\end{split}
\end{equation}

\paragraph{Heterogeneous Cooperation}

For heterogeneous cooperation, we consider four different combinations as follow:
\begin{itemize}
    \item (\textit{x Carriers}, \textit{y Suppliers}), $x + y = 3m$;
    \item (\textit{x Carriers}, \textit{z Observers}), $x + z = 3m$;
    \item (\textit{y Suppliers}, \textit{z Observers}), $y + z = 3m$;
    \item (\textit{x Carriers}, \textit{y Suppliers}, \textit{z Observers}), $x + y + z = 3m$;
\end{itemize}

Then, we estimate the expected amount of rescued agents $U_e$ and energy cost $E_e$ for each group.

\textit{a. (\textit{x Carriers}, \textit{y Suppliers})}

In this scenario, we consider \textit{Carrier} and \textit{Supplier} have the similar sensing range, and they both have enough energy (\textit{Basic Needs}) to support the entire rescue mission. So we can present $\mathbb{E}(U_{e1})$ and $\mathbb{E}(E_{e1})$ as Eq. \eqref{re1} and \eqref{ee1}.
\begin{equation}
\nonumber
\begin{split}
    \mathop{\mathbb{E}(U_{e1})} = & \frac{t_n(1 - e^{-\lambda_{e1}})}{\lambda_{e1}} \times ((x \times cap_c + y \times cap_s) \cap \\
    & (x \times res_c + y \times res_s)),
\end{split}
\end{equation}
\begin{equation}
    \lambda_{e1} = \frac{2l}{v_c} + \frac{2t_ccn}{3m \times sen_c}
\label{re1}
\end{equation}
\begin{equation}
\begin{split}
    \mathop{\mathbb{E}(E_{e1})} = & e_t + \frac{2cn}{3m \times sen_c}e_c + \\
    & \frac{t_n(1 - e^{-\lambda_{e1}})}{\lambda_{e1}} \times ((x \times cap_c + y \times cap_s) \cap \\
    & (x \times res_c + y \times res_s))
\label{ee1}
\end{split}
\end{equation}

\textit{b. (\textit{x Carriers}, \textit{z Observers})}

Here, we assume the entire group's velocity adapt \textit{Carriers'} speed. Similarly, we can express $\mathbb{E}(U_{e2})$ and $\mathbb{E}(E_{e2})$ as Eq. \eqref{re2} and \eqref{ee2},
\begin{equation}
\nonumber
\begin{split}
    \mathop{\mathbb{E}(U_{e2})} = & \frac{t_n(1 - e^{-\lambda_{e2}})}{\lambda_{e2}} \times ((x \times cap_c + z \times cap_o) \cap \\
    & (x \times res_c + z \times res_o)),
\end{split}
\end{equation}
\begin{equation}
    \lambda_{e2} = \frac{2l}{v_c} + \frac{2t_ccn}{x \times sen_c + z \times sen_o}
\label{re2}
\end{equation}
\begin{equation}
\begin{split}
    \mathop{\mathbb{E}(E_{e2})} = & e_t + \frac{2cn}{x \times sen_c + z \times sen_o}e_c + \\
    & \frac{t_n(1 - e^{-\lambda_{e2}})}{\lambda_{e2}} \times ((x \times cap_c + z \times cap_o) \cap \\
    & (x \times res_c + z \times res_o))
\label{ee2}
\end{split}
\end{equation}

\textit{c. (\textit{y Suppliers}, \textit{z Observers})}

$\mathbb{E}(U_{e3})$ and $\mathbb{E}(E_{e3})$ as Eq. \eqref{re3} and \eqref{ee3},
\begin{equation}
\nonumber
\begin{split}
    \mathop{\mathbb{E}(U_{e3})} = & \frac{t_n(1 - e^{-\lambda_{e3}})}{\lambda_{e3}} \times ((y \times cap_s + z \times cap_o) \cap \\
    & (y \times res_s + z \times res_o)),
\end{split}
\end{equation}
\begin{equation}
    \lambda_{e3} = \frac{2l}{v_c} + \frac{2t_ccn}{y \times sen_s + z \times sen_o}
\label{re3}
\end{equation}
\begin{equation}
\begin{split}
    \mathop{\mathbb{E}(E_{e3})} = & e_t + \frac{2cn}{y \times sen_s + z \times sen_o}e_c + \\
    & \frac{t_n(1 - e^{-\lambda_{e3}})}{\lambda_{e3}} \times ((y \times cap_s + z \times cap_o) \cap \\
    & (y \times res_s + z \times res_o))
\label{ee3}
\end{split}
\end{equation}

\textit{d. (\textit{x Carriers}, \textit{y Suppliers}, \textit{z Observers})}

$\mathbb{E}(U_{e4})$ and $\mathbb{E}(E_{e4})$ as Eq. \eqref{re4} and \eqref{ee4}.
\begin{equation}
\nonumber
\begin{split}
    \mathop{\mathbb{E}(U_{e4})} = & \frac{t_n(1 - e^{-\lambda_{e4}})}{\lambda_{e4}} \times \\
    & ((x \times cap_c + y \times cap_s + z \times cap_o) \cap \\
    & (x \times res_c + y \times res_s + z \times res_o)),
\end{split}
\end{equation}
\begin{equation}
    \lambda_{e4} = \frac{2l}{v_c} + \frac{2t_ccn}{x \times sen_c + y \times sen_s + z \times sen_o}
\label{re4}
\end{equation}
\begin{equation}
\begin{split}
    \mathop{\mathbb{E}(E_{e4})} = & e_t + \frac{2cn}{x \times sen_c + y \times sen_s + z \times sen_o}e_c + \\
    & \frac{t_n(1 - e^{-\lambda_{e4}})}{\lambda_{e4}} \times \\
    & ((x \times cap_c + y \times cap_s + z \times cap_o) \cap \\
    & (x \times res_c + y \times res_s + z \times res_o))
\label{ee4}
\end{split}
\end{equation}

\subsection{Comparative Analysis}

After above discussion, in this section, we first compare the performances between (\textit{Homogeneous vs Homogeneous}), (\textit{Heterogeneous vs Heterogeneous}) and (\textit{Homogeneous vs Heterogeneous}), then analyze the exiting of optimal or suboptimal solution for heterogeneous cooperation system in rescue mission. In order to simplify calculation, we assume Eq. \eqref{assume} and also regard \textit{Carrier} and \textit{Supplier} have the similar sensing range, and the sensing range of group \textit{Observer} approaches infinity.
\begin{equation}
\begin{split}
    res_c = cap_s = cap_o = res_o = k,~~k \in Z^+
\label{assume}
\end{split}
\end{equation}

\textit{a. \textit{Homogeneous vs Homogeneous}}

Comparing the expected amount of rescued agents between those groups can be represented as Eq. \eqref{hr_com}.
\begin{equation}
\begin{split}
    \mathop{\mathbb{E}(U_{hc}) : \mathbb{E}(U_{hs}) : \mathbb{E}(U_{ho})} = 1 : 1 : \frac{\lambda_{hc}(1 - e^{-\lambda_{ho}})}{\lambda_{ho}(1 - e^{-\lambda_{hc}})}
\label{hr_com}
\end{split}
\end{equation}

Also, we can compare the group expectation energy cost of \textit{Carrier} and \textit{Supplier}, \textit{Carrier} and \textit{Observer} and \textit{Supplier} and  \textit{Observer} as Eq. \eqref{he_cs_com} and \eqref{he_cso_com} respectively.
\begin{equation}
\begin{split}
    \mathop{\mathbb{E}(E_{hc}) - \mathbb{E}(E_{hs})} = 0
\label{he_cs_com}
\end{split}
\end{equation}
\begin{equation}
\begin{split}
    \mathop{\mathbb{E}(E_{hc}) - \mathbb{E}(E_{ho})} & = \mathbb{E}(E_{hs}) - \mathbb{E}(E_{ho}) = \\
    & \frac{2cn}{3m \times sen_{c}}e_c + 3mkt_n(\frac{1 - e^{-\lambda_{hc}}}{\lambda_{hc}} \\
    & - \frac{1 - e^{-\lambda_{ho}}}{\lambda_{ho}})
\label{he_cso_com}
\end{split}
\end{equation}

Through the above discussion, if we assume that \textit{Observers} also does not concern about their energy cost (\textit{Basic Needs}) in the entire rescue mission, they will have the best performance comparing with other groups. Actually, in reality, the energy level and consumption rate of \textit{Observer}, like the drone, are much lower and faster than \textit{Carrier} and \textit{Supplier} correspondingly, which means that \textit{Observer} need to waste lots of time to charge. Considering this issue, we assume that these three groups have a similar performance generally to simplify our calculation.

\textit{b. \textit{Heterogeneous vs Heterogeneous}}

Similarly, considering involving \textit{Observers} in the group, the entire group sensing range approach infinity. And according to the assumption Eq. \eqref{com}, \eqref{sen}, \eqref{res} and \eqref{cap}, we can estimate the heterogeneous comparison of the expectation amount of rescued agents as Eq. \eqref{er_com}.
\begin{equation}
\begin{split}
    \mathop{\mathbb{E}(U_{e1})} :~& \mathbb{E}(U_{e2}) : \mathbb{E}(U_{e3}) : \mathbb{E}(U_{e4}) \approx \\ 
    & \frac{\lambda_{e0}(1 - e^{-\lambda_{e1}})}{\lambda_{e1}(1 - e^{-\lambda_{e0}})} : \frac{x \times res_c + z \times res_o}{x \times cap_c \cap y \times res_s} : \\
    & \frac{y \times cap_s + z \times cap_o}{x \times cap_c \cap y \times res_s} : 1,~~~\lambda_{e0} = \frac{2l}{v_c}
\label{er_com}
\end{split}
\end{equation}

The corresponding group expected energy cost comparison is shown as Eq. \eqref{ee_12_com}, \eqref{ee_23_com} and \eqref{ee_24_com}.
\begin{equation}
\begin{split}
    \mathop{\mathbb{E}(E_{e1}) - \mathbb{E}(E_{e2})} \approx & \frac{2cn}{3m \times sen_{c}}e_c + \\
    & t_n((x \times cap_c \cap y \times res_s)\frac{1 - e^{-\lambda_{e1}}}{\lambda_{e1}} \\
    & - 3mk\frac{1 - e^{-\lambda_{e0}}}{\lambda_{e0}}) > 0
\label{ee_12_com}
\end{split}
\end{equation}
\begin{equation}
\begin{split}
    \mathop{\mathbb{E}(E_{e2}) - \mathbb{E}(E_{e3})} = 0
\label{ee_23_com}
\end{split}
\end{equation}
\begin{equation}
\begin{split}
    \mathop{\mathbb{E}(E_{e2}) - \mathbb{E}(E_{e4})} \approx & t_n\frac{1 - e^{-\lambda_{e0}}}{\lambda_{e0}}(3mk - \\
    & (x \times cap_c \cap y \times res_s)) > 0
\label{ee_24_com}
\end{split}
\end{equation}

According to Eq. \eqref{er_com}, \eqref{ee_12_com}, \eqref{ee_23_com} and \eqref{ee_24_com}, we can notice that the performance of the low bound and the high bound in those groups are the combination of \textit{(Carrier \& Supplier)} and \textit{(Carrier \& Supplier \& Observer)} respectively.

\textit{c. \textit{Homogeneous vs Heterogeneous}}

As the above discussion, at this stage, we compare the performance between low bound of heterogeneous cooperation system and homogeneous cooperation system as Eq. \eqref{ehr_com} and \eqref{ehe_com}.
\begin{equation}
\begin{split}
    \mathop{\mathbb{E}(U_{e1}) : \mathbb{E}(U_{hc})} \approx \frac{x \times cap_c \cap y \times res_s}{3m \times res_c} > 1
\label{ehr_com}
\end{split}
\end{equation}
\begin{equation}
\begin{split}
    \mathop{\mathbb{E}(E_{e1}) - \mathbb{E}(E_{hc})} \approx & t_n\frac{1 - e^{-\lambda_{e1}}}{\lambda_{e1}}((x \times cap_c \cap y \times res_s) \\
    & - 3mk)) < 0
\label{ehe_com}
\end{split}
\end{equation}

According to Eq. \eqref{ehr_com}, we can notice that the \textit{Expectation Utility} of heterogeneous cooperation system is larger than the homogeneous cooperation system. Also, Eq. \eqref{ehe_com} shows that the homogeneous cooperation system's energy cost is higher than the heterogeneous system.

\begin{figure*}[tbp]
\centering
\includegraphics[width=1\textwidth]{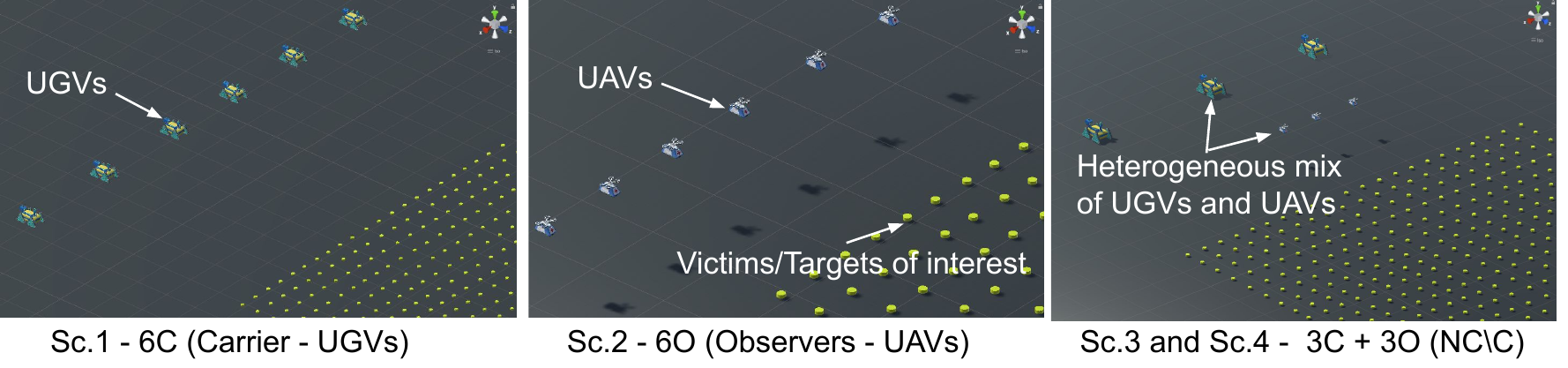}
\caption{Illustration of the four scenarios with homogeneous and heterogeneous team of \textit{Carrier} (UGV) and \textit{Observer} (UAV) in a rescue mission simulation. Scenario 3 is non-cooperative (NC) between the UGVs and UAVs and Sc. 4 is cooperative (C) between the different type of robots.}
\label{fig: rescuing_scenarios}
\end{figure*}
 
 \begin{figure}[tbp]
\centering
\includegraphics[width=0.5\textwidth]{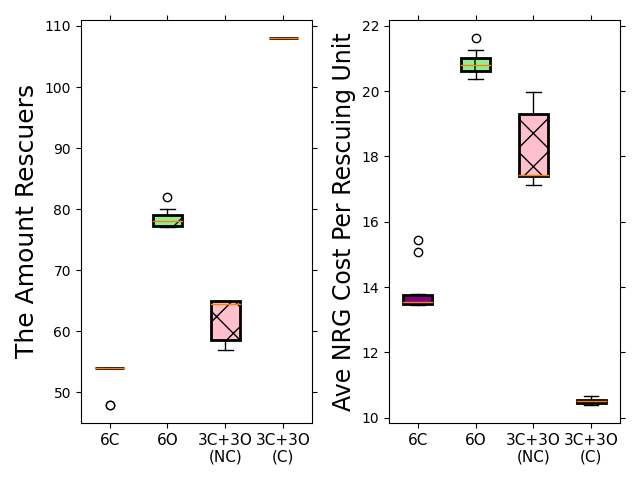}
\caption{The analysis of experiments' results on homogeneous and heterogeneous MRS cooperation in simulation.}
\label{fig: data_merge}
\end{figure}

\section{Numerical Evaluation}

To simulate the above problem, we use "Unity" game engine and build a simple scenario (see Fig. \ref{fig: rescuing_scenarios}) to verify our results. We design two kinds of experiments -- \textit{Homogeneous} and \textit{Heterogeneous} MRS Cooperation and consider two categories of robots -- \textit{Carrier} and \textit{Observer} implemented in the specific experiments. Video demonstration of the experiments is available at \url{http://hero.uga.edu/research/heterogeneous-cooperation/}.

We suppose the common category has the same battery level in the initial state, and in every moving step, carrier and observer will cost 0.045\% and 0.015\% energy separately. To simplify the group utility's visualization, we do not consider any obstacles, rescue resource requirement Eq. \eqref{res}, and communication energy cost. We design four scenarios -- homogeneous part simulates six carriers (Car) and six observers (Obs) fulfilling rescue mission correspondingly and considering three carriers and three observers cooperation (C) and non-cooperation (NC) for heterogeneous MRS. We also implement a simple \textit{Negotiation-Agreement Mechanism} \cite{yang2020hierarchical, mrs2019} to avoid the collision in the whole process.
 
To compare the performance of \textit{Homogeneous} and \textit{Heterogeneous} MRS in the experiments, we calculate the amount rescuers (Group Utility) and the average energy cost per rescuing unit in five minutes Eq. \eqref{rescue_mission_problem}. Considering observer limited energy store (basic needs) Eq. \eqref{eng}, we assume that if the individual energy level is below 30\%, it needs to go to rest place charging 10 seconds, then back to work. Also, we assume the observer can perceive the whole map. In the homogeneous scenarios, due to working in an uncertain environment with limited perception range, the carrier's velocity is equal to a tenth of the observer's speed for avoiding uncertainty risks and satisfying its safety needs Eq. \eqref{vel}. But with the observers' assist in heterogeneous MRS cooperation, carriers can share information with observers, enlarge their perception range and double their velocity. And observers will decrease half of the speed to adapt carriers' involvement. Each carrier and observer can rescue eight and one units respectively in each round Eq. \eqref{cap}. For a non-cooperation heterogeneous system, the two groups do not interact and fulfill the mission separately.
  
According to the above assumption, we conduct ten simulation trials for each scenario. Fig. \ref{fig: data_merge} shows the number of rescuers and average energy cost per rescuing unit, respectively. For the homogeneous MRS cooperation, comparing with the performance of group carrier and observer separately, although observer can achieve higher group utility (the amount rescuers) than carrier \ref{fig: data_merge}(a1) in a limited time Eq. \eqref{hr_com}, the average energy cost per rescuing unit represents more consumption \ref{fig: data_merge}(a2). On the other hand, for the heterogeneous MRS, the non-cooperation system shows a medial performance comparing with the different three scenarios, which does not offer distinguished advantages. Generally speaking,  the heterogeneous MRS cooperation not only delivers more excellent group utility Eq. \eqref{ehr_com} and less system cost Eq. \eqref{ehe_com} from the system perspective, but also saves more cost per rescuing unit from the individual angle. 
 
More importantly, from the statistical perspective (Fig. \ref{fig: data_merge}), comparing with the rest of the scenarios, the heterogeneous cooperation system decreases performance uncertainty (deviation between trials) and provides more stability and robustness for the whole system. It can help the system adapting more complex and uncertain environments efficiently and presents more robust viability.

\section{Application to Human-Robot Teaming}
\label{sec:humanrobotteaming}

As the higher-level intelligent creature globally, humans represent more complex and diversified needs such as personal security, health, friendship, love, respect, recognition, and so forth. When we consider humans and robots work as a team, organizing their needs and getting a common ground is a precondition for human-robot collaboration in urban search and rescue missions.

From a robot needs perspective, it first needs to guarantee human security and health, such as avoiding collision with humans, protecting them from radiation, and so forth. But in the higher level teaming needs, robots should consider human team members' specialty and capability to form corresponding heterogeneous \textit{Human-Robot} team adapting specific rescue missions automatically.

Humans also expect robots to provide safety and a stable working environment in aiding rescue missions from human needs. Furthermore, efficient and reliable assistance plays an essential element for the entire rescue missions. More importantly, designing an \textit{Interruption Mechanism} can help humans interrupt robots' current actions and re-organize them to fulfill specific emergency tasks or execute some crucial operations manually.

The individual robot learning model can be regarded as constructing models of the other agents, which takes as input some portion of the observed interaction history, and returns a prediction of some property of interest regarding the modeled agent \cite{albrecht2018autonomous}. In our future work, we enable robots to learn and adapt to human needs and keep up trust and rapport between humans and robots, which are critical for the task efficiency and safety improvement \cite{nourbakhsh2005human}. 
Here, the adaptive learning of \textit{Human-Robot Interaction} will be pursued along the following lines:
\begin{itemize}
    \item Adopting suitable formation to perceive and survey environments predicting threats (and warn human team members) and explore new rescue tasks.
    \item Reasonable path planning adaptation in various scenarios avoid collision guaranteeing human security and decreasing human working environment interference.
    \item Combining the specific capabilities and needs of robots and humans, calculating sensible strategies to organize the entire group collaboration fulfilling corresponding rescue mission efficiently.
\end{itemize}

Using the above line of thought, assume we can model the human needs and find a way to calculate the expected utility of the human member in a human-robot team, then the proposed framework can be applied to a human-robot team by integrating the human needs with robot needs and capabilities specified in Sec.~\ref{sec:problem}. It is expected to result in a harmonious teaming with heterogeneous agents (humans/robots) in achieving common goals.

\section{Conclusion}

We presented an overview of the needs-driven cooperation model for heterogeneous multi-robot systems and theoretically analyzed the importance of heterogeneity in increasing rescue mission performance. We advanced the robot needs hierarchy established in our earlier work, formalized the general rescue mission, and categorized the robots in USAR missions as carriers, suppliers, and observers.

We theoretically evaluated the system's performance in terms of the group utility and energy cost to achieve the rescue mission in a limited time. We proved that the needs-drive cooperation in a heterogeneous robot system enabled higher group utility than a homogeneous robot system. We also demonstrated the advantages of needs-driven heterogeneous cooperation through simulation experiments involving two groups of robots, namely carriers (UGVs)  and observers (UAVs) in our experiment design. The results verified that heterogeneous multi-robot cooperation increased group utility and robustness and decreased energy costs and performance uncertainties compared to the homogeneous multi-robot grouping for the same task execution. 

Future work will focus on extending this work to human-robot teaming and how the system as a whole can enable self-learning at the robot-level.

\bibliographystyle{IEEEtran}
\bibliography{references}



\end{document}